# Spin-polarized quasiparticle transport in cuprate superconductors

C.-C. Fu, Z. Huang, and N.-C. Yeh

*Department of Physics, California Institute of Technology, Pasadena, California 91125*

The effects of spin-polarized quasiparticle transport in superconducting $YBa_2Cu_3O_{7-\delta}$ (YBCO) epitaxial films are investigated by means of current injection into perovskite ferromagnet-insulator-superconductor (F-I-S) heterostructures. These effects are compared with the injection of simple quasiparticles into control samples of perovskite nonmagnetic metal-insulator-superconductor (N-I-S). Systematic studies of the critical current density ($J_c$) as a function of the injection current density ($J_{inj}$), temperature ($T$), and the thickness ($d$) of the superconductor reveal drastic differences between the F-I-S and N-I-S heterostructures, with strong suppression of $J_c$ and a rapidly increasing characteristic transport length near the superconducting transition temperature $T_c$ only in the F-I-S samples. The temperature dependence of the efficiency ($\eta \equiv \Delta J_c / J_{inj}$; $\Delta J_c$: the suppression of critical current due to finite $J_{inj}$) in the F-I-S samples is also in sharp contrast to that in the N-I-S samples, suggesting significant redistribution of quasiparticles in F-I-S due to the longer lifetime of spin-polarized quasiparticles. Application of conventional theory for nonequilibrium superconductivity to these data further reveal that a substantial chemical potential shift $\mu^*$ in F-I-S samples must be invoked to account for the experimental observation, whereas no discernible chemical potential shift exists in the N-I-S samples, suggesting strong effects of spin-polarized quasiparticles on cuprate superconductivity. The characteristic times estimated from our studies are suggestive of anisotropic spin relaxation processes, possibly with spin-orbit interaction dominating the $c$-axis spin transport and exchange interaction prevailing within the $CuO_2$ planes. Several alternative scenarios attempted to account for the suppression of critical currents in F-I-S samples are also critically examined, and are found to be neither compatible with experimental data nor with the established theory of nonequilibrium superconductivity.

## I. INTRODUCTION

One of the most intriguing open questions associated with high-temperature superconductivity is the relevance of $d_{x^2-y^2}$-wave pairing symmetry[1,2] and antiferromagnetic spin correlation[3] to the pairing mechanism, and the possibility of spin-charge separation due to either the resonant-valence-bond (RVB) scenario[4–6] or the existence of charged stripes.[7,8] A natural consequence of the $d_{x^2-y^2}$-wave pairing symmetry in the hole-doped ($p$-type) cuprate superconductors[1,2] is an anisotropic pairing potential and the existence of nodal quasiparticles that are responsible for the unconventional low-energy excitation spectra.[9,10] The doping of holes into the oxygen $2p$ orbitals in the $CuO_2$ planes is known to incur spin fluctuations in the cuprates due to the frustration of nearest-neighbor antiferromagnetic $Cu^{2+}$-$Cu^{2+}$ correlation, and the existence of spin fluctuations has been suggested as relevant to the $d_{x^2-y^2}$-wave pairing in the cuprates.[3] The antiferromagnetic correlation has also motivated the RVB scenario that could lead to spin-charge separation and the marginal Fermi-liquid (MFL) electronic behavior[11] in the normal state. However, to date there has been no direct evidence for spin-charge separation in the cuprates.

The existence of $d_{x^2-y^2}$-wave pairing and antiferromagnetic correlation is also believed to be responsible for the unusual response of $p$-type cuprates to quantum impurities.[12–16] That is, the substitution of $Cu^{2+}$ by nonmagnetic impurities (such as $Li^+$, $Zn^{2+}$, $Mg^{2+}$, and $Al^{3+}$) in the $CuO_2$ planes of $Bi_2Sr_2CaCu_2O_x$ and $YBa_2Cu_3O_{7-\delta}$ (YBCO) systems has revealed stronger pair-breaking effects than the magnetic impurities (such as $Ni^{2+}$),[17–28] in sharp contrast to the insensitivity of conventional superconductors to nonmagnetic impurities.[29,30] In light of the unconventional response to static magnetic and nonmagnetic impurities and in search of possible evidence for spin-charge separation in the cuprates,[31] a feasible experimental approach is to compare the spin and charge transport in the cuprate superconductors. Such investigation may be conducted by comparing the cuprate response to externally injected simple and spin-polarized quasiparticles, and the physical description for the experimental phenomena would involve concepts of nonequilibrium superconductivity.[32]

Nonequilibrium superconductivity and its associated phenomena have been studied extensively since the 1970s,[33] and the primary focus of the research has been on the effects of either simple (i.e., spin-degenerate) quasiparticle injection[34–38] or photon-induced Cooper-pair breaking and quasiparticle redistribution[39–45] in conventional $s$-wave superconductors. The nonequilibrium effects have yielded observation of both enhancement[39–43] and suppression[34–38,45] of superconductivity. In the rarely studied case of injection of spin-polarized quasiparticles, two primary effects on the suppression of superconductivity must be considered.[46] One is associated with the resulting excess magnetic moments that break the time-reversal symmetry in singlet superconductors.[47] The other is the excess momentum and the nonequilibrium quasiparticle distribution that modify the energy spectrum of the superconductor.[32,33] In the absence of significant scattering by either magnetic impurities or spin–orbit coupling, the transport lifetime of spin-polarized quasiparticles is expected to be much longer than that of simple quasiparticles due to the low probability of pair recombination with other quasiparticles.[46] However, the complexity of the combined effects aforementioned has limited theoretical development at the microscopic level for spin-polarized quasiparticle transport in superconductors.

Spin injection into superconductors can be accomplished by passing electrical currents through a ferromagnet before the tunneling across a thin insulating barrier into a superconductor.[48,49] In recent years, the injection of spin-polarized current in perovskite ferromagnet–insulator–superconductor (F-I-S) heterostructures has attracted significant experimental interest.[50–54] This technique utilizes the excellent lattice match among various perovskite materials for epitaxial film growth[52] of the heterostructures, and also takes advantage of the half-metallic ferromagnetism of perovskite manganites[55–59] that yield much better spin polarization than typical metallic ferromagnets. Thus, investigating the characteristic spin and charge relaxation and transport processes in the perovskite F-I-S and N-I-S devices can be a unique vehicle for probing nonequilibrium superconductivity and possibly the pairing mechanism in the cuprates. Indeed, strong suppression of the superconducting critical current has been observed in cuprate superconductors by injecting currents from the underlying half-metallic ferromagnetic manganite films.[50–53,60] In our recent publication,[52,53] possible complications due to Joule heating incurred from large injection currents through resistive insulator and ferromagnet layers were minimized by employing a pulsed current technique.[52,53] The resulting experimental data reveal insignificant effects of simple quasiparticle injection in the control samples of perovskite non-magnetic metal-insulator-superconductor (N-I-S) heterostructures, whereas F-I-S samples with comparable geometry exhibit strong suppression of critical currents and significant modification to the quasiparticle density of states (DOS).[53,54] Consequently, the experimental findings are attributed to the dynamic pair-breaking effects of spin-polarized quasiparticles as a result of excess magnetic moments and quasiparticle redistribution.[52–54]

Despite a significant number of experimental reports that are supportive of the effects of spin-injection in cuprates, many important issues are yet to be resolved. Experimentally, determining the magnitude and temperature dependence of the spin-relaxation length and time has proven to be elusive. Theoretically, microscopic interactions of externally injected spin-polarized quasiparticles with the Cooper pairs and existing quasiparticles in cuprate superconductors remain unknown. Nonetheless, the intrinsic anisotropy in the cuprate superconducting order parameter due to the predominant $d_{x^2-y^2}$-wave pairing symmetry[28,61–63] and the weakly interacting-layered structure[64] are expected to be relevant to the spin and charge transport. For instance, the in-plane simple quasiparticle recombination time $\tau_R$ in DyBa$_2$Cu$_3$O$_{7-\delta}$ is found to be significantly longer ($\tau_R \approx 10^{-6} - 10^{-5}$ s) than the typical values ($\tau_R \approx 10^{-9} - 10^{-7}$ s) in conventional superconductors. This phenomenon is attributed to the tendency of simple quasiparticles relaxing towards the zeros of the superconducting gap and also to the reduced scattering rate of nodal quasiparticles by phonons.[65] In addition to the interaction with nodal quasiparticles, quasistatic injection of spin-polarized quasiparticles into the cuprates can suppress the antiferromagnetic correlation in the CuO$_2$ planes, which may result in significant and long-range effects on the cuprate superconductivity, similar to the strong influence of nonmagnetic quantum impurities in the CuO$_2$ planes.[16]

In this report, we extend our previous studies of nonequilibrium superconductivity by quantifying various characteristics of spin injection in F-I-S with a range of thickness for the superconducting layer. Studies of N-I-S partner heterostructures are also included as controls. By comparing the degree of critical current suppression $\Delta J_c$ in the presence of external injection at different YBCO thickness, we are able to infer a rapidly increasing c-axis spin relaxation length near $T_c$ in F-I-S, while no such divergence exists in the N-I-S samples. Furthermore, an empirically defined efficiency ($\eta$, which measures the suppression of critical currents due to injected quasiparticles), is studied in detail for both F-I-S and N-I-S systems. We find that the efficiency in F-I-S is strongly dependent on temperature and exhibits nonmonotonic dependence on the injection current density ($J_{inj}$). In contrast, the efficiency in N-I-S is smaller than that in F-I-S for all temperatures and is monotonic with $J_{inj}$. These results suggest that spin-polarized quasiparticles exert strong effects on suppressing the cuprate superconductivity, probably due to their strong influence on the short-range Cu$^{2+}$-Cu$^{2+}$ antiferromagnetic coupling and the intimate correlation of superconductivity with the background antiferromagnetism. We also critically examine several alternative scenarios attempting to account for the experimental findings without invoking the effects of spin injection, and find that these alternative scenarios are neither compatible with empirical facts nor consistent with any established theory of nonequilibrium superconductivity. Finally, we remark that our work is primarily concerned with the spin and charge transport properties inside the superconducting cuprates after quasiparticle transmission across the interfaces of the heterostructures. For in-depth consideration of quasiparticle transport across the interface of unconventional superconductors with various types of metals and for different crystalline axes, the readers may refer to other theoretical studies[66–71] and experimental investigation.[72,73]

This paper is structured as follows. In Sec. II the sample fabrication and characterization together with the experimental methods are described. The results derived from our experiments are given in Sec. III, with detailed analysis presented in Sec. IV. A critical examination of several alternative scenarios is given in Sec. V. Finally, Sec. VI summarizes our findings and the physical implications of the results.

## II. EXPERIMENT

The c-axis-oriented trilayer F(N)-I-S heterostructures used in this work contained YBa$_2$Cu$_3$O$_{7-\delta}$ (YBCO) as the superconductor, SrTiO$_3$ (STO) as the insulator, La$_{0.7}$Sr$_{0.3}$MnO$_3$ (LSMO) or La$_{0.7}$Ca$_{0.3}$MnO$_3$ (LCMO) as the ferromagnet, and LaNiO$_3$ (LNO) as the nonmagnetic metal. A number of devices were studied with different thicknesses of the constituent layers, and for the F-I-S devices, the choice of either LSMO or LCMO did not yield any discernible differences.[52] The thickness of YBCO ranged from 40 nm to 160 nm, of LSMO or LCMO and of LNO was kept constant at 100 nm, and of STO was either 2 nm or 3.5 nm. The samples were fabricated using the pulsed laser deposition technique on (6 mm×6 mm) LaAlO$_3$ (LAO) substrates, with either LSMO or LNO as the lower layer and YBCO as the top layer, and the insulator buffering in between. Details of the fabrication condition have been given elsewhere.[57,58] The close lattice match among the constituent layers of the perovskite F-I-S and the substrates facilitated epitaxial film growth,[50] thus minimizing strong spin–flip scattering at the interface and preserving spin polarization during injection. For electrical contact, each of the YBCO and STO (LNO) layers had four gold pads placed on top using sputtering deposition. The compositional quality of these heterostructures were examined using x-ray photoelectron spectroscopy (XPS).[52] To ensure no discernible reaction between layers during the growth process, XPS studies of bilayers of YBCO/STO and STO/LSMO on LAO were monitored and the absence of reaction within ~0.1 atomic percent for at least the top 10 nm of the YBCO layer was confirmed.

To further verify the quality of samples, electrical transport measurements were performed on both the superconducting and ferromagnetic layers to determine the normal-state resistivity $\rho_n$ and the transition temperatures $T_c$ and $T_{Curie}$. In addition, scanning tunneling spectroscopy was also performed on the YBCO layer of the F-I-S and N-I-S samples, and the superconducting gap value was found to be consistent with that of the optimally doped YBCO single crystals.[53,54] The LAO substrate was chosen because it had been demonstrated to be the substrate that yielded minimum lattice strain and the best magnetization alignment for the thin-film growth of manganites.[57,58,74] Our characterizations revealed that the resistivity of each constituent layer of the heterostructures and the Curie temperature ($T_{Curie}$) of the ferromagnetic manganites were all comparable to those of the corresponding single crystalline materials.[52] Since the resistivity of the manganite is known to couple strongly to the magnetic properties and therefore is a characterization for the quality of the manganite, a manganite layer with resistivity comparable to that of a single crystal implies large and relatively well aligned ferromagnetic domains.[57–59,74]. We also note that the temperature dependence of the resistivity in the manganite layer always exhibited either a maximum or a distinct change in slope near $T_{Curie}$ (~260 K for LCMO and ~320 K for LSMO) which was characteristic of high-quality ferromagnetic manganites.[57,58,75] On the other hand, the superconducting transition temperature ($T_c$) of YBCO varied somewhat among devices, ranging from 84 to 90 K, with no apparent correlation with the YBCO thickness. We attribute the $T_c$ variation to uncertainties in the substrate temperature during the thin film growth. Due to the variation in $T_c$, the temperature dependence of various physical quantities of YBCO shall be considered in reduced temperature ($T/T_c$) rather than absolute temperature $T$.

The critical current ($I_c$) measurements of the YBCO were made with the pulsed current technique, which synchronized two pulsed current generators that supplied a measurement current through YBCO and an injection current through the metallic underlayer, reference to a common ground, as illustrated in Fig. 1(a). The advantage of this method was to eliminate undesired Joule heating on the YBCO from power dissipation in the event when high current levels (<300 mA) passed through the electrical contacts and the resistive metallic underlayer. A 1:1000 ratio of the pulsed current width ($t_w$) to period ($t_p$) was chosen, which yielded a negligible temperature increase (<10 mK) in the YBCO during maximum current injection, monitored with *in situ* thermometry using the resistivity of the manganite. The pulse width used for this work was $t_w = 300$ μs. Also shown in Fig. 1(a) is a schematic illustration of the dimension of the as-grown heterostructures and the positions of the electrodes. The lateral dimension of the LSMO or LNO was (6×6) mm$^2$, and that of the YBCO layer was (6×2) mm$^2$.

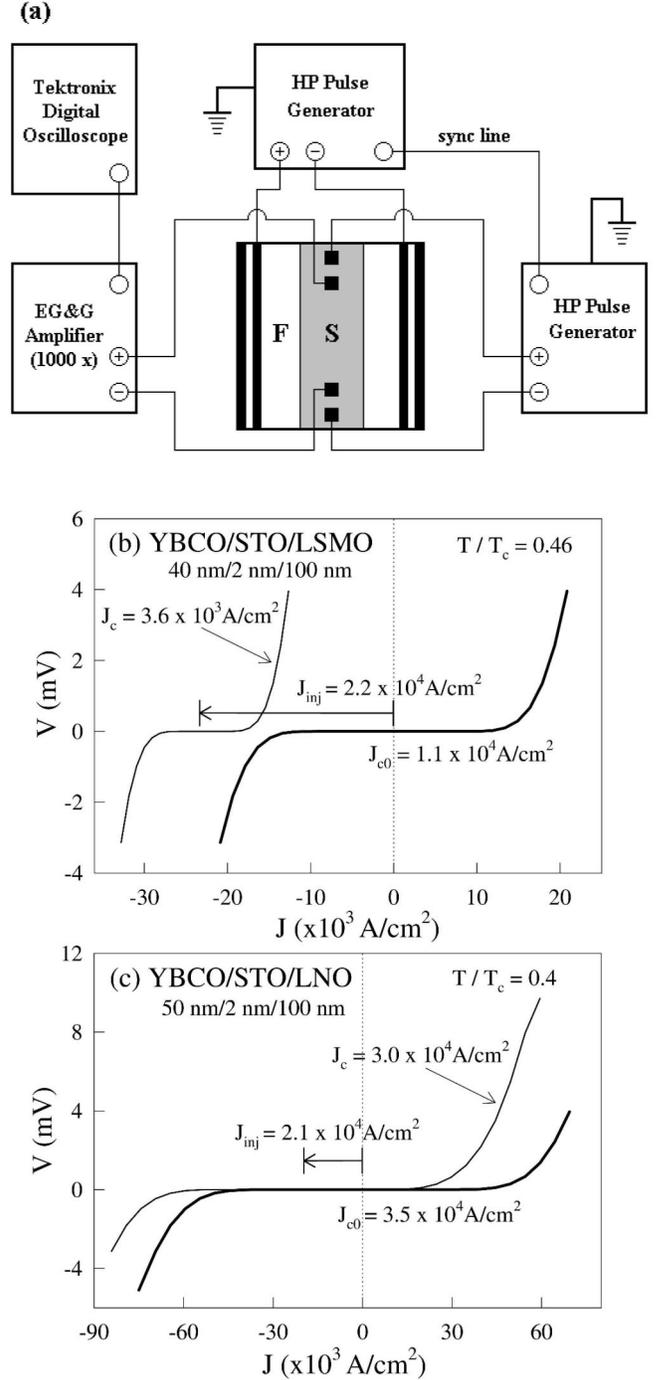

FIG. 1. (a) Block diagram of the pulsed-current measurement setup. (b) Representative current-voltage ($I$-$V$) characteristics of an F-I-S sample at ($T/T_c$)=0.46, showing a significant left shift of the $I$-$V$ curve and a substantial suppression of $J_c$ upon injection of currents from a ferromagnetic layer. (c) Representative $I$-$V$ characteristics of an N-I-S sample at ($T/T_c$)=0.4, showing a much smaller left shift of the $I$-$V$ curve and much weaker suppression of $J_c$ than those in the F-I-S samples upon injection of comparable currents from a nonmagnetic layer.

In addition to the as-grown heterostructures described above, we also attempted measurements on two sets of F-I-S devices with much smaller lateral dimensions for the YBCO layer, (including 100×100 μm$^2$, 10×100 μm$^2$, 5×100 μm$^2$, 2×100 μm$^2$, and 1×100 μm$^2$), which were patterned using photolithography and ion milling techniques. The first set of F-I-S devices were made on YBCO/STO/LSMO of thicknesses 100 nm/2 nm/100 nm. The second set were on similar samples of thicknesses 100 nm/3.5 nm/150 nm. We found that the normal-state YBCO layer of the patterned F-I-S samples generally exhibited larger resistivity, by a factor of 2 to 3, than those of single crystals and as-grown heterostructures, suggesting some deterioration of the bulk sample quality after device processing. In particular, among the various lateral widths of YBCO in the patterned F-I-S devices, no superconducting transition was observed for the 2 μm and 1 μm devices, although they were electrically continuous. It is therefore reasonable to infer that the damage to the edge of the YBCO layer due to the patterning process extended over a width on the order of a few microns. This damaged region was comparable to the experimentally estimated transfer length of ~1.8 μm for the first set of F-I-S and ~3.5 μm for the second set, where the transfer length[76–78] is a measure of the characteristic region in length that the injected current transfers from the underlying layer into the YBCO superconductor. Thus, the degree of spin polarization for the injected currents in the patterned F-I-S devices might have been much weakened because of strong magnetic impurity scattering within the damaged region at the interface. Furthermore, the YBCO layer of the patterned F-I-S devices exhibited a gradual degradation in both $T_c$ and $J_{c0}$ after each thermal cycling, together with sporadic superficial discoloration after large external current injection. Similar attempts on patterning N-I-S devices yielded even worse results, with severe degradation to the samples so that the YBCO layer was either only superconducting below 20 K or not superconducting at all down to 4.2 K. Hence, it is difficult to draw reliable conclusions from data taken on these patterned devices, pending further improvements on device processing to achieve better sample quality and robustness.

In Fig. 1(b), generic *I-V* curves of an F-I-S heterostructure with zero and a finite injection current are shown. The curve symmetric about the zero-current axis corresponds to the *I-V* data in the presence of no current injection. For a given temperature, we define the current values that drive the YBCO superconductor to register $+3$ μV and $-3$ μV across its voltage terminals (~3 mm apart) as the critical currents $I_c^+$ and $I_c^-$, respectively. The second curve to the left shows a shifted *I-V* curve because of an external injection current $I_{inj}$ that increases the total current passing through the superconductor. This effect is present for the injection of both simple and spin-polarized quasiparticles. The observed narrowing of the gap in between the $I_c^+$ and $I_c^-$ values with increasing $I_{inj}$ is the result of critical current suppression due the apparent deterioration of superconductivity from the external perturbation. Therefore, the critical current under quasiparticle injection is defined as $I_c = (I_c^+ - I_c^-)/2$, and that in the absence of quasiparticle injection is $I_{c0}(T)$. The magnitude of the shift in the *I-V* is related to the amount of current entering the superconductor from the underlayer current injection and is hereafter defined as $I_{inj}$. Such shifts are always present under external injection, as exemplified in Figs. 1(b) and 1(c) for F-I-S and N-I-S samples with comparable thicknesses of YBCO and similar reduced temperatures. The critical current density $J_c$ and the injection current density $J_{inj}$ are obtained by dividing the corresponding currents by the cross section of the superconductor. We note in Figs. 1(b) and 1(c) that suppression in $J_c$ is much more significant in the F-I-S sample. Additional *I-V* curves for F-I-S heterostructures at other reduced temperatures and for a range of injection currents are shown in Figs. 2(a) and 2(b) for further comparison.

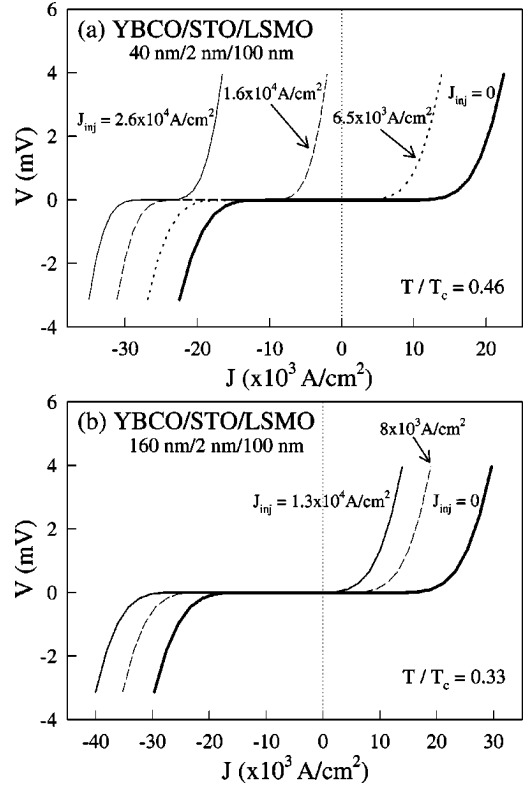

FIG. 2. Representative *I-V* characteristics of F-I-S samples with YBCO thickness of (a) 40 nm at $(T/T_c) = 0.46$ and (b) 160 nm at $(T/T_c) = 0.33$ for a range of injection currents. This illustrates the significant $J_c$ suppression observed in the thin YBCO heterostructure (a).

Besides the effect of injection currents on $J_c$, we have also reported previously[52] that the low-temperature critical current density $J_{c0}$ in the absence of injection is sensitive to the thickness of the insulator barrier of the F-I-S heterostructures, with systematically increasing $J_{c0}$ for samples with thicker insulating barriers and otherwise identical lateral dimensions. Similar finding has also been confirmed in our patterned F-I-S devices. Furthermore, the $J_{c0}$ values of N-I-S samples at low temperatures were larger than the corresponding $J_{c0}$ of F-I-S samples with the same lateral dimensions and barrier thickness. Such a systematic dependence has ruled out the possibility that self-field induced edge-vortex dissipation might have been the primary cause of $J_{c0}$ suppression with decreasing insulating barrier, and has been attributed to a "self-injection" phenomenon.[52,79]

It is worth noting that the pulsed-current setup employed in our experiment involved the use of pulsed-voltage generators, which linked sources with output impedances comparable to the relevant resistance in the measurement circuit. This setup therefore resulted in a finite, but small, leakage current flow through the pulse generator upon the introduction of injection current from the underlayer. However, simple circuit analysis and direct calibration had indicated that the leakage current was less than 10% of the total injected current for all measurements. In principle, decoupled current paths can be achieved with smaller and lithographically defined devices and with the use of high output-impedance current sources. Indeed we have made and studied several F-I-S devices with smaller lateral dimensions, ranging from 100 $\mu$m to 1 $\mu$m. However, due to the aforementioned issues with sample quality and edge damage, experimental results obtained on these patterned F-I-S devices were not conclusive. Hence, we shall concentrate on the experimental studies of the larger as-grown devices in this report, and only return to some of the results obtained on the patterned devices in Sec. V to address issues concerning alternative hypotheses for the suppression of critical currents in F-I-S heterostructures.

A useful definition for experimental characterization of our devices, in normalized current densities, is given by

$$\eta(T, J_{inj}) \equiv \frac{\Delta J_c(T, J_{inj})}{J_{inj}} \equiv \frac{[J_{c0}(T) - J_c(T, J_{inj})]}{J_{inj}}, \quad (1)$$

where $\eta$ is defined as the efficiency of quasiparticle injection that relates the magnitude of critical current suppression to a given amount of injection current. The temperature and injection current dependence of the efficiency for the F-I-S and N-I-S heterostructures with the same YBCO thickness can provide insightful comparison for the spin and charge transport in the cuprate superconductors.

In addition to the dependence of critical currents on $J_{inj}$ and ($T/T_c$), we have investigated F-I-S and N-I-S samples with difference thicknesses of YBCO in order to deduce viable information for a characteristic spin relaxation length ($\delta_s$). A number of F-I-S devices with different YBCO thicknesses ($d = 40$ nm, 50 nm, 100 nm, and 160 nm) together with their corresponding N-I-S control samples ($d = 50$ nm and 100 nm) have been fabricated and studied. We note that the $J_{c0}$ values at 4.2 K were not a monotonic function of $d$, with $J_{c0} = 5.8 \times 10^4$ A/cm$^2$ for $d = 40$ nm, $5.2 \times 10^4$ A/cm$^2$ for $d = 50$ nm, $7.0 \times 10^4$ A/cm$^2$ for $d = 100$ nm, and $1.5 \times 10^4$ A/cm$^2$ for $d = 160$ nm. Detailed current-injection effects on these F-I-S and N-I-S samples are described in the following section.

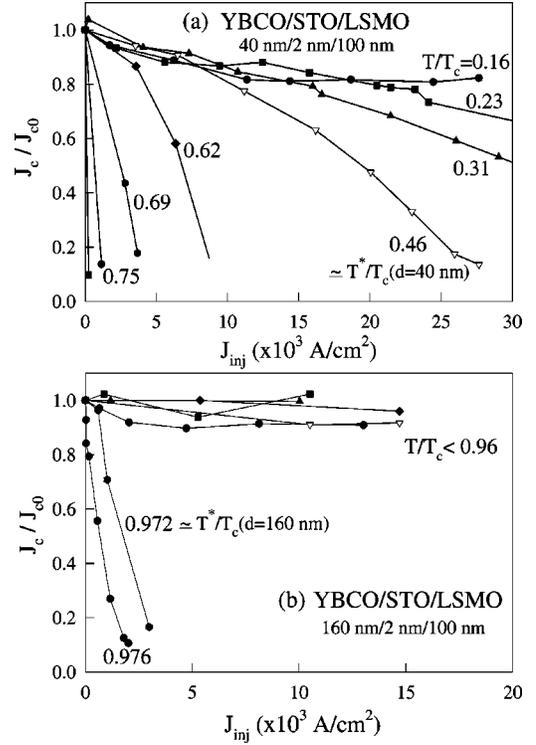

FIG. 3. $J_c$-vs-$J_{inj}$ isotherms of (a) an F-I-S sample with YBCO thickness $d = 40$ nm and $J_{c0} = 5.8 \times 10^4$ A/cm$^2$ at 4.2 K; (b) an F-I-S sample with YBCO thickness $d = 160$ nm and $J_{c0} = 1.5 \times 10^4$ A/cm$^2$ at 4.2 K.

### III. RESULTS

The critical current density ($J_c$) provides a macroscopic measure that empirically characterizes the effect of quasiparticle injection on superconductivity. Given a constant thickness of the insulating barrier and the same lateral dimensions of the superconductor, $J_c$ is determined by the temperature ($T$), the injection current density ($J_{inj}$), the characteristic sample dimension ($d$) along the direction of quasiparticle injection, and the microscopic mechanism for quasiparticle transport across the interface and interaction in the superconductor. The dependence of $J_c$ on the YBCO thickness is the result of a finite quasiparticle relaxation length along the $c$ axis of the superconductor if all other parameters are kept the same. Through this dependence, we can estimate the $c$-axis spin-polarized and simple quasiparticle relaxation lengths by studying F-I-S and N-I-S with a range of different YBCO thickness. Two sets of representative $J_c$-vs-$J_{inj}$ isotherms taken on F-I-S heterostructures with $d = 40$ and 160 nm are shown in Figs. 3(a) and 3(b), respectively. We found that nearly full suppression of critical current could be achieved at lower reduced temperatures in the F-I-S with a thinner (40 nm) YBCO than that with a thicker (160 nm) YBCO F-I-S heterostructure, with the latter only beginning to exhibit discernible suppression due to current injection above the reduced temperature $>0.97$. This result is consistent with the notion of a finite $c$-axis spin-polarized quasiparticle relaxation length. That is, the manifestation of nearly complete critical current suppression should correlate closely with a $c$-axis spin-relaxation length $\delta_s^c(T)$ approaching the YBCO thickness ($d$). Thus, we expect spin-polarized quasiparticles to survive throughout nearly the entire thickness of YBCO when strong $J_c$ suppression is observed. Under this premise, studies of the $J_c$-vs-$J_{inj}$ isotherms for F-I-S samples with different YBCO thicknesses can provide a viable measure for the temperature dependence of the $c$-axis spin-relaxation length. In contrast, the relative ratio of critical current suppression by a finite $J_{inj}$ at a given ($T/T_c$) was appreciably smaller in the N-I-S samples, as shown in Refs. 52 and 53 where no discernible $J_c$ suppression could be detected in an N-I-S sample with a YBCO thickness $d=100$ nm.

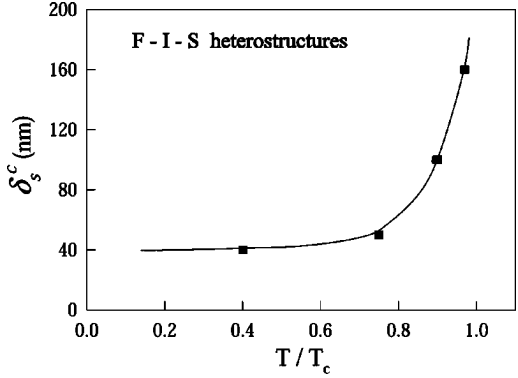

FIG. 4. Characteristic $c$-axis spin-relaxation length $\delta_s^c$ as a function of reduced temperature ($T/T_c$) for the F-I-S heterostructures.

To estimate the $c$-axis spin-relaxation length $\delta_s^c(T)$ in YBCO, we empirically related the YBCO thickness $d$ of each F-I-S heterostructure to a characteristic reduced temperature $[T^*(d)/T_c]$ at which $(J_c/J_{c0})\leq 0.1$ is satisfied under a constant $J_{inj}$. This assignment was based on the assumption that the observation of strong suppression in $J_c$ corresponded to the condition $\delta_s^c \to d$ for $(T/T_c)\to[T^*(d)/T_c]$, provided that the lateral dimensions of all samples were kept identical. Similar criterion could be applied to the N-I-S samples to define the $c$-axis charge relaxation length $\delta_Q^c$. The correlation of $(T^*/T_c)$ with the corresponding thickness ($d$) of the F-I-S heterostructure is shown in Fig. 4, suggesting a characteristic length $\delta_s^c$ $[\sim d$ for $(T/T_c)\to(T^*/T_c)]$ increased rapidly near $T_c$. The diverging characteristic length was attributed to a vanishing superconducting gap near $T_c$,[32] and was only detectable in the F-I-S heterostructures. We further remark that the diverging behavior in F-I-S samples was unlikely the result of any extrinsic effect such as systematically varying quality of YBCO with its thickness for the following reasons. All F-I-S samples had comparable $T_c$ while their $J_{c0}$ values at 4.2 K were not monotonic with increasing film thickness, with minimum $J_{c0}\sim 1.5\times 10^4$ A/cm$^2$ associated with the sample of maximum thickness $d=160$ nm. Furthermore, in contrast to the observation in F-I-S samples, no obvious crossover temperature $T^*$ could be found in N-I-S samples for rapid decrease of $J_c$ with $J_{inj}$. Thus, the occurrence of strong injection-induced superconductivity suppression (with $J_c<\sim 0.1 J_{c0}$) at larger values of $(T^*/T_c)$ for thicker F-I-S samples could not be ascribed to the result of better superconductivity in thicker YBCO.

The contrast in the temperature dependence of $\delta_s^c$ and of $\delta_Q^c$ for the F-I-S and N-I-S samples could be attributed to the significantly longer lifetime of spin-polarized quasiparticles relative to that of simple quasiparticles,[33,46] so that the injection of simple quasiparticles did not result in complete suppression of $J_c$ in N-I-S samples for all temperatures of our studies. In other words, the condition $J_c(T,J_{inj})<\sim 0.1 J_{c0}(T)$ could not be realized in the N-I-S samples within our experimental resolution,[52,53] so that the charge relaxation length $\delta_Q^c$ appeared to be always smaller than $d$ for all temperatures of our study. We shall discuss this phenomenon and the contrast between F-I-S and N-I-S samples more quantitatively in Sec. IV.

In addition to the strong dependence of the $J_c$-vs-$J_{inj}$ behavior on the thickness of F-I-S heterostructures, we also compared the efficiencies $\eta(T,J_{inj})$ in F-I-S and N-I-S samples, which were considered to better quantify the suppression of $J_c$ due to nonequilibrium quasiparticle injection. Our definition of the efficiency in Eq. (1) is equivalent to the definition of a "gain" in the devices by others.[50,51] As shown in Figs. 5(a) and 5(b), a distinct contrast was observed between the isotherms of the efficiency in F-I-S ($\eta_s$) and those in N-I-S ($\eta_n$) devices as a function of $J_{inj}$. In general, $\eta_s$ in F-I-S was significantly larger than the corresponding $\eta_n$ in N-I-S for all reduced temperatures. Evidently, an anomalous strong decrease in $\eta_s$ with increasing $J_{inj}$ was found only in F-I-S samples at low temperatures. Furthermore, for reduced temperatures $0.16\leq(T/T_c)\leq 0.31$ in F-I-S, $\eta_s$ exhibited a nonmonotonic dependence with $J_{inj}$, and then became monotonically increasing with $J_{inj}$ for $0.31<(T/T_c)<1$. Interestingly, we note that at low spin-polarized quasiparticle injections, the "gain" was actually greater than unity. In contrast, $\eta_n$ for the control N-I-S devices appeared to increase monotonically with $J_{inj}$ at all temperatures, and the magnitude of $\eta_n$ was always much smaller than unity.

## IV. ANALYSIS

The seemingly surprising contrast between the F-I-S and N-I-S samples may be understood in the context of different quasiparticle relaxation mechanisms and nonequilibrium quasiparticle distributions. Generally speaking, an adequate description for nonequilibrium superconductivity must involve consideration of the quasiparticle energy $E_\mathbf{k}$ and the quasiparticle distribution function $f_\mathbf{k}$ in the superconductor, where $\mathbf{k}$ denotes the quasiparticle momentum. In principle, an explicit expression for $E_\mathbf{k}$ can be obtained by solving the Bogoliubov–de Gennes equations,[32,80] provided that the exact Hamiltonian $\mathcal{H}$ for the superconductor is known. In thermodynamic equilibrium, the quasiparticle energy associated with the unperturbed Hamiltonian $\mathcal{H}_0$ is $E_{\mathbf{k}0}=(\Delta_{\mathbf{k}}^2+\xi_{\mathbf{k}}^2)^{1/2}$, where $\xi_{\mathbf{k}}(\equiv\varepsilon_{\mathbf{k}}-\varepsilon_F)$ is the single particle energy $\varepsilon_{\mathbf{k}}$ relative to the Fermi level $\varepsilon_F$, and $\Delta_{\mathbf{k}}$ is the pairing potential.[32,80] For an $s$-wave superconductor $\Delta_{\mathbf{k}}$ is a constant, whereas for a pure $d_{x^2-y^2}$-pairing superconductor, $\Delta_{\mathbf{k}}\approx\Delta_d\cos 2\theta_k$, and $\theta_k$ is an angle measured from one of the antinodes of the order parameter in momentum space.[1,2] The injection of external quasiparticles is expected to interact with the superconductor through an interaction Hamiltonian $\mathcal{H}_I$ and to modify the quasiparticle energy and the distribution of quasiparticle states through the total Hamiltonian $\mathcal{H}=\mathcal{H}_0+\mathcal{H}_I$, provided that the perturbative approximation is valid. In the absence of available theory for nonequilibrium quasiparticle distributions in a strongly correlated $d$-wave superconductor, we consider in the following analyses of our data based on conventional theory of nonequilibrium superconductivity, and discuss the implication of results thus derived from F-I-S and N-I-S samples.

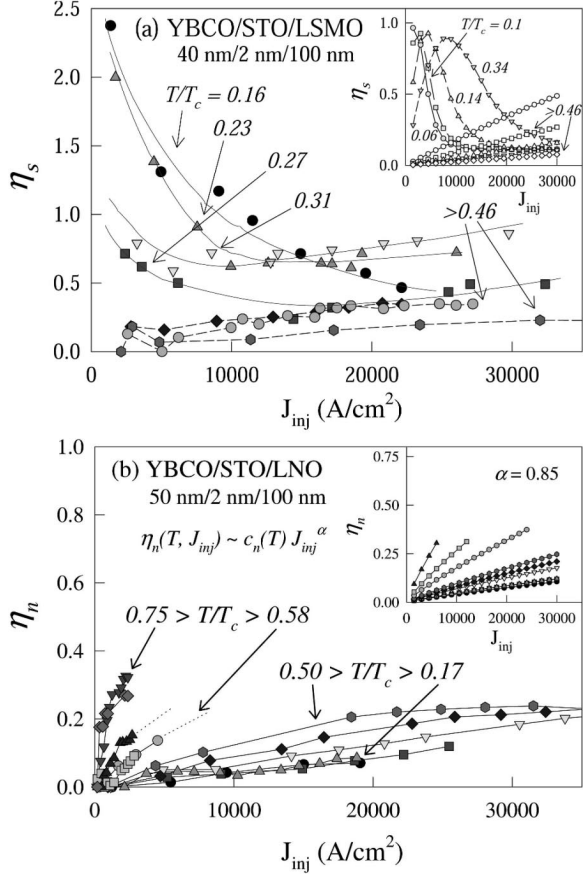

FIG. 5. (a) Efficiency $\eta_s$-vs-$J_{inj}$ isotherms of an F-I-S sample with YBCO thickness $d=40$ nm. Inset: Simulated results using Eq. (8). (b) Efficiency $\eta_n$-vs-$J_{inj}$ isotherms of an N-I-S sample with YBCO thickness $d=50$ nm. Inset: Simulated results using Eq. (3).

Consider a simple case where a uniform supercurrent with a velocity $\vec{v}_s=\vec{J}_s/(n_s e)$ exists in the superconductor. The finite momentum associated with the supercurrent $J_s$ is found to change the quasiparticle energy $E_{\mathbf{k}n}$ and the distribution function $f_{\mathbf{k}n}$ as follows:[32]

TABLE I. The values and references of various physical parameters used in computing $\eta_s$ and $D_s^c$ are tabulated. Here $t$ and $t'$ are the nearest- and next-nearest-neighbor interaction integrals in the two-dimensional tight-binding model for the normal state energy band structure $\varepsilon_{\mathbf{k}}$ of the CuO$_2$ plane (Ref. 81).

| Parameter | Value | Comment |
| --- | --- | --- |
| $\Delta_d$ | 30 meV | Eqs. (2), (4) |
| $\varepsilon_F$ | 0.51 eV | Ref. 81 |
| $t$ | 0.18 eV | Eq. (4); Ref. 81 |
| $t'$ | 50 meV | Eq. (4); Ref. 81 |
| $\gamma$ | 1.02 | Eq. (4); Ref. 81 |
| $v_F$ | $2\times 10^5$ m/s$^2$ | Eq. (10) |
| $l_{tr}^c$ | $\sim 1$ nm | Eq. (10) |

$$E_{\mathbf{k}n}=E_{\mathbf{k}0}+\hbar\mathbf{k}_F\cdot\mathbf{v}_s/m^*\equiv E_{\mathbf{k}0}+\delta E_J,$$

$$f_{\mathbf{k}n}=\{1+\exp[E_{\mathbf{k}n}/(k_B T)]\}^{-1}, \quad (2)$$

where $\mathbf{k}_F$ is the quasiparticle momentum at the Fermi level. Using the $d_{x^2-y^2}$-wave pairing potential in optimally doped YBCO with $\Delta_d\approx 30$ meV,[61–63] and the single particle energies $\varepsilon_{\mathbf{k}}$ derived from the tight-binding band structure calculations[81] with parameters tabulated in Table I, we find that $E_{\mathbf{k}0}\gg|\delta E_J|$ is satisfied for typical supercurrents ($J_c=10^4$–$10^5$ A/cm$^2$ for $T\ll T_c$) sustainable in the YBCO superconducting layers. Thus, we expect $f_{\mathbf{k}n}\sim 0$.

In the event that a quasiparticle current $J_{inj}$ is externally injected into a superconductor that already carries a supercurrent, the situation becomes more complex because the externally injected quasiparticles must redistribute themselves among available states that obey the Pauli exclusion principle for fermions, and the redistribution must involve inelastic and elastic scattering processes. Therefore the injected quasiparticle momenta relative to the supercurrent direction and the lattice momenta are not well defined due to the involvement of scattering processes,[32] and it is not uncommon that a current-carrying superconductor with an initial supercurrent $J_s$ can remain superconducting under an external injection current $J_{inj}$ such that the sum of $J_{inj}$ and $J_s$ exceeds the critical current $J_{c0}$ of the superconductor,[32] as exemplified in Figs. 1(b) and 2(a). Nonetheless, we find that the simple sum of the maximum injection current density and the supercurrent density still yields $\max[\hbar\vec{k}_F\cdot(\vec{J}_s+\vec{J}_{inj})/(n_s e)]\ll E_{\mathbf{k}0}$ at low temperatures. Thus, the excess momentum due to external quasiparticle injection is insufficient to yield significant redistribution of quasiparticles. For N-I-S samples with relatively thin YBCO, the small yet monotonically increasing $\Delta J_c$ with increasing $J_{inj}$ and increasing $T$ [see Fig. 5(b)] should be attributed primarily to the increasing normal fluid density and the suppression of the superconducting phase stiffness.

More specifically, the efficiency associated with simple quasiparticle injection in the N-I-S samples may be given by the following phenomenological expression:

$$\eta_n\equiv\frac{1}{N_0}\sum_k(1-2f_{\mathbf{k}n})c_n(T)g(J_{inj})\approx c_n(T)g(J_{inj}), \quad (3)$$

where $N_0$ denotes the total number of quasiparticle states, $c_n(T)$ is associated with the temperature-dependent fraction of the normal fluid and is a monotonic function of $T$, $g(J_{inj})$ reflects the weakening of the superconducting phase stiffness under increasing $J_{inj}$ and is a monotonic function of $J_{inj}$, and the quantity $(1-2f_{kn})$ ensures no double occupancy of the quasiparticle states. However, we have found $f_{kn} \approx 0$ for the entire range of $T$ and $J_{inj}$ of our interest, so that $N_0^{-1}\Sigma_k(1-2f_{kn}) \approx 1$. Empirically, we find that $c_n(T) \sim \{1-[1-(T/T_c)]^\nu\}$ where $0.5 < \nu < 0.9$ and $g(J_{inj}) \sim (J_{inj})^{0.85}$. The simulated example of $\eta_n$-vs-$J_{inj}$ isotherms using Eq. (3) and the values $\nu \approx 0.6$ and $\alpha \approx 0.85$ are shown in the inset of Fig. 5(b) for comparison with the experimental data.

In contrast, significantly different response of the cuprate superconductors to the injection of spin-polarized quasiparticles is expected because of their relatively longer lifetime and their strong effects on suppressing the $Cu^{2+}$–$Cu^{2+}$ antiferromagnetic correlation. Although we do not know the exact interaction Hamiltonian for the spin-polarized quasiparticles in the cuprate superconductors and therefore cannot obtain the quasiparticle energy $E_{ks}$, it seems informative to estimate the approximate quasiparticle energy by applying conventional theory for nonequilibrium quasiparticle distribution under charged particle injection to the cuprate superconductors. That is, we assume the validity of perturbative approximation as manifested by an effective chemical potential shift $\mu^*$ in the single particle energy $\xi_\mathbf{k}$.[32,34,37] Noting that $J_c$ is obtained by identifying the onset of dissipation where the maximum magnitude of the $d_{x^2-y^2}$-wave superconducting gap has been driven to a small value, we consider the situation similar to that for a gapless superconductor.[80] Thus, the quasiparticle energy under spin injection may be approximated by

$$E_{\mathbf{k}s}=[(\xi_\mathbf{k}-\mu^*)^2+\Delta_\mathbf{k}^2]^{1/2} \approx |\xi_\mathbf{k}-\mu^*|\left[1+\frac{\Delta_\mathbf{k}^2}{2(\xi_\mathbf{k}-\mu^*)^2}\right], \quad (4)$$

where $\Delta_k \ll |\xi_\mathbf{k}-\mu^*|$, and the corresponding quasiparticle distribution function becomes

$$f_{\mathbf{k}s}(T,J_{inj})=1/\{1+\exp[E_{\mathbf{k}s}/(k_B T)]\}. \quad (5)$$

The chemical potential shift per quasiparticle $\mu^*$ due to spin injection must satisfy the conditions $\mu^* \to 0$ for $J_{inj} \to 0$ and $\mu^* \to$ constant $\equiv \mu_{00}^*$ for large $J_{inj}$ and low temperatures, where $\mu_{00}^*$ is a constant. Therefore a reasonable approximation for $\mu^*$ can be given by $\mu^* = \mu_{00}^* \tanh(c_1 J_{inj}/k_B T)$, and the physical significance of the functional form $\tanh(c_1 J_{inj}/k_B T)$ is consistent with the average spin polarization ($P$) per quasiparticle in the superconductor. Here $c_1 J_{inj} = \mu_B[\mu_0(\langle m_s \rangle P/\Omega)]$ is associated with the effective field energy, $\langle m_s \rangle$ denotes the excess magnetic moments in the superconductor due to spin injection, $\mu_B$ is the Bohr magneton, $\mu_0$ is the vacuum permeability, and $\Omega = \mathcal{A}d$ is the superconducting volume. In the absence of a known interaction Hamiltonian, $\mu_{00}^*$ and $c_1$ are positive quantities to be determined empirically. On the other hand, simple dimensional analysis yields

$$\langle m_s \rangle = (\mu_B/e)I_{inj}\tau_s = (\mu_B/e)(J_{inj}\mathcal{A}\tau_s), \quad (6)$$

where $\tau_s$ is the spin-dephasing time.

Anticipating suppression in the critical current density due to the presence of excess magnetic moments, we may related $\Delta J_c$ to the effective magnetization ($\langle m_s \rangle P/\Omega$) adjusted by the available quasiparticle states. That is,

$$\Delta J_c \equiv (J_{c0}-J_c) \propto \sum_k (1-2f_{ks})(\langle m_s \rangle P/\Omega)$$

$$\propto \sum_k (1-2f_{ks})J_{inj}\tanh(c_1 J_{inj}/k_B T), \quad (7)$$

where the effective magnetization ($\langle m_s \rangle P/\Omega$) takes the form of the Brillouin function for a spin-1/2 system, and the quantity $(1-2f_{ks})$ ensures no double occupancy in the quasiparticle states. As a result, $\eta_s$ becomes

$$\eta_s \equiv \Delta J_c/J_{inj} \sim \frac{1}{N_0}\sum_k (1-2f_{ks})\tanh(c_1 J_{inj}/k_B T). \quad (8)$$

Strictly speaking, the quantity $(1-2f_{ks})$ in Eq. (8) should have been written as $(1-f_\uparrow-f_\downarrow)$, where the $\uparrow$ spin polarization corresponds to that parallel to the effective magnetic field induced by the excess magnetic moments. However, the effective field can be shown to be very small, as discussed in Sec. V. Consequently, we find $(1-f_\uparrow-f_\downarrow) \approx (1-2f_{ks})$.

Following the analysis outlined above and inserting the relevant experimental parameters as tabulated in Table I into Eq. (8), we obtained results similar to the experimental findings for the F-I-S sample with the thinnest YBCO in which the effects of spin injection were fully realized over a wide temperature range, as shown in the inset of Fig. 5(a). However, a quantitative agreement with the experimental data for the thinnest F-I-S sample could only be achieved by invoking a large chemical potential shift $\mu_{00}^*$ associated with $f_{ks}$, so that $\mu^*$ varied from $\sim 700$ meV at $T \ll T_c$ to $\sim 45$ meV at $T \to T_c$. These values are unusually large, comparable to the band structure parameters. Interestingly, the empirical value $\mu_{00}^*$ for $T \ll T_c$ is comparable to $4J_{ex}$ in the YBCO system, where $J_{ex}$ is the nearest-neighbor antiferromagnetic coupling constant, and the factor of 4 corresponds to the number of nearest neighbors in the square lattice of the $CuO_2$ plane. While these large values of $\mu_{00}^*(T)$ are likely unphysical and should not be taken literally because of the questionable validity of applying conventional theory to cuprate superconductivity, the following conclusions may be drawn from our analyses. First, the large magnitude of $\mu_{00}^*$ found only in F-I-S implies strong effects of spin injection on cuprate superconductors, as opposed to the negligible change in the chemical potential of N-I-S samples under current injection. Second, the large $\mu_{00}^*$ values in F-I-S suggest the breakdown of conventional perturbative approximation to the interaction Hamiltonian of nonequilibrium superconductors. That is, a valid perturbative approximation would have yielded a small chemical potential shift relative to the relevant band structure parameters in the single-particle energy $\xi_\mathbf{k}$. That the chemical potential shift derived from perturba-tive approximation turned out to be comparable to the band structure parameters is suggestive of strong interaction effects associated with spin-polarized quasiparticles in cuprate superconductors.

Despite the uncertainty in the magnitude of the chemical potential, the temperature dependence of $\mu^*$ is directly related to that of the effective magnetization, and therefore can provide information for the spin-relaxation process. In particular, for the F-I-S sample with the thinnest YBCO layer ($d=40$ nm), the spin-injection effects were already realized at low temperatures, suggesting that the $c$-axis spin-relaxation length was either comparable to or exceeding the sample thickness over most temperatures of our investigation. Hence, the temperature evolution of the spin-dependent information deduced from those data may be considered to be primarily associated with that of the in-plane spin relaxation. In contrast, measurements on F-I-S samples with thicker YBCO contained convoluted information for both the $c$-axis and in-plane spin-relaxation processes over most temperatures except near $T_c$, and therefore could not be used to infer direct information associated with the in-plane spin relaxation. Thus, the empirically determined coefficient $c_1(T)$ in Eq. (8) for the F-I-S sample with $d=40$ nm could be approximately related to an effective in-plane spin-relaxation time $\tau_s(T)$ by $c_1(T) \approx \mu_0 \mu_B^2 \tau_s/(ed)$, and we find that $\tau_s(T)$ ranges from $\sim 10^{-4}$ s at $T \ll T_c$ to $\sim 10^{-6}$ s at $T \to T_c^-$. Such a long characteristic time scale is comparable to the spin-spin relaxation time obtained from the nuclear quadruple resonance (NQR) experiments,[82] and is approximately one-to-two orders of magnitude longer than the in-plane simple quasiparticle recombination time determined from measurements of photoinduced activation of microwave absorption.[65]

The above phenomenological analyses suggest that the injection of spin-polarized quasiparticles in YBCO appeared to exert strong influence on the microscopic quasiparticle energy and density of states (DOS), probably through exchange interaction with the short-range $Cu^{2+}$–$Cu^{2+}$ antiferromagnetic correlation. Furthermore, the slower relaxation of spin-polarized quasiparticles relative to the already long recombination time of simple quasiparticles[65] appeared reasonable because of the further reduced probability of quasiparticle recombination before excess spin polarization can be relaxed. It is also interesting to compare the transport data presented here with our scanning tunneling spectroscopic studies of YBCO in the F-I-S and N-I-S heterostructures that revealed significantly modified quasiparticle DOS at 4.2 K under spin injection and no discernible changes under simple quasiparticle injection.[54] The spectroscopic studies are not only supportive for our finding of significantly longer relaxation time of spin-polarized quasiparticles relative to that of simple quasiparticles, but also suggestive of direct influence of spin injection on the microscopic states of the cuprates.

Next, we consider the appearance of a diverging spin-relaxation length near $T_c$. In conventional superconductors, it is known that the characteristic quasiparticle relaxation time $\tau_Q$ can diverge near $T_c$ due to the vanishing superconducting gap $\Delta(T)$ through the following relation[32,33]

$$\tau_Q(T) \approx \frac{4\tau_E k_B T_c}{\pi \Delta(T)}, \qquad (9)$$

where $\tau_E$ is the inelastic electron-phonon scattering time, and $\Delta(T) = \Delta_0 [1-(T/T_c)]^\nu$, with $\Delta_0$ being the zero-temperature superconducting gap and $\nu$ the order-parameter critical exponent. This diverging behavior gives rise to stronger effects of quasiparticle injection with increasing temperature near $T_c$. The temperature interval for revealing such divergence depends on the critical fluctuation regime and also on the temperature dependence of $\tau_E$, and is generally very narrow in conventional superconductors, because $\tau_Q$ decreases rapidly with decreasing temperature and competes with other characteristic times (such as the quasiparticle recombination time) at low temperatures. On the other hand, the critical fluctuation regime of cuprate superconductors is known to be several orders of magnitude larger than that of the conventional superconductors.[83] In the case of YBCO, the critical regime associated with the zero-field transition temperature $T_c$ is estimated at approximately 1%–10% of $T_c$. Hence, it is in principle more promising to observe this diverging quasiparticle relaxation length in the cuprates near $T_c$.

In the preceding section, we have attributed the rapidly increasing characteristic length near $T_c$ in the F-I-S samples (see Fig. 4) to the $c$-axis spin-relaxation length $\delta_s^c$. Whereas the transport of spin-polarized quasiparticles actually took place along both in-plane and $c$ axis, this attribution is still reasonable because the $c$-axis dimensions of all F-I-S samples were several orders of magnitude smaller than the lateral dimensions and therefore were most sensitive to the crossover of a $c$-axis relaxation length to the sample thickness. Consequently, the temperature dependence of $\delta_s^c$ could be related to a $c$-axis spin-relaxation time $\tau_s^c$, at least semiquantitatively. If the spin transport along the $c$ axis is diffusive and if no spin-charge separation exists, we have $\delta_s^c = \sqrt{D_s^c \tau_s^c}$, where $D_s^c = (v_F \ell_{tr}^c)/(3\lambda_{so})$ is the $c$-axis spin-diffusion coefficient and $\ell_{tr}^c$ is the transport mean-free path along the $c$ axis, $\lambda_{so}(\sim 0.1)$ is the dimensionless spin–orbit coupling constant,[31] and $\tau_s^c$ is associated with the inelastic spin–orbit scattering time $\tau_{so}$ via a relation similar to that in Eq. (9):

$$\tau_s^c(T) \approx \frac{4\tau_{so} k_B T_c}{\pi \langle |\Delta_{\mathbf{k}}(T)| \rangle_k}. \qquad (10)$$

Here $\langle |\Delta_{\mathbf{k}}(T)| \rangle_{\mathbf{k}}$ denotes the angular average of the $d$-wave gap, and the temperature dependence of $|\Delta_{\mathbf{k}}(T)|$ is approximated by $|\Delta_{\mathbf{k}}(T)| \sim [1-(T/T_c)]^\nu$. Assuming that $\tau_{so}$ is a weak function of the temperature, we compare the $\delta_s^c$ value ($\sim 40$ nm) at $(T/T_c)=0.1$ with that ($\sim 160$ nm) at $(T/T_c)=0.9$ and use Eq. (10) to obtain the order-parameter exponent $\nu \approx 0.65$, which is consistent with the exponent $\nu=2/3$ for the $XY$ model. Furthermore, using Eq. (10), the physical parameters $v_F$ and $\ell_{tr}^c$ listed in Table I, and the empirical values of $\delta_s^c(T)$, we find $\tau_{so} \approx 10^{-11} - 10^{-10}$ s, which is consistent with the theoretical estimate for spin-orbit interaction.[31] We therefore suggest that $c$-axis spin-relaxation mechanism may be dominated by the spin-orbit interaction, and the relaxation time $\tau_s^c$ is substantially shorter than that associated with the in-plane spin relaxation, implying anisotropic spin transport.

Concerning the $c$-axis simple quasiparticle transport, we remark that the overall effects of current injection in the N-I-S samples depend strongly on the transmission and energy relaxation of simple quasiparticles along the $c$ axis, and are therefore sensitive to the inter-planar inelastic scattering mechanism in addition to the in-plane quasiparticle recombination. Given that the $c$-axis dimensions of the N-I-S samples were much smaller than the lateral dimensions, the overall effects of simple quasiparticle injection should be primarily determined by the magnitude of the $c$-axis simple quasiparticle relaxation length $\delta_n^c$ relative to the sample thickness, even though the in-plane recombination time of excess simple quasiparticles can be relatively long due to the existence of nodes in the pairing potential.[65] Taking Eq. (9) and the typical electron-phonon scattering time in the cuprates, $\tau_E \sim 10^{-11}$ s for $(T/T_c) \ll 1$ and $\tau_E \sim 10^{-13}$ s for $(T/T_c) \to 1$, we obtained $\delta_n^c = \sqrt{D_n^c \tau_Q}$ that ranges from $\sim 20$ nm at $(T/T_c) \sim 0.1$ to $< \sim 5$ nm at $(T/T_c) \sim 0.9$, where $D_n^c = v_F l_{tr}^c/3$ is the charge diffusion coefficient along the $c$-axis. These estimates are consistent with the negligible effect of current injection in the N-I-S samples with a thick superconducting layer ($\sim 100$ nm), and the finite (although relatively small) suppression of $J_c$ in those N-I-S samples with a thin superconducting layer ($\sim 50$ nm). Due to the rapid decrease in the electron-phonon scattering time $\tau_E$ with $T$ near $T_c$, a diverging $\delta_n^c(T)$ can only be expected if temperature becomes sufficiently close to $T_c$ so that the increasing value of $\langle \Delta(T) \rangle^{-1}$ with $T$ compensates for the decreasing $\tau_E(T)$. A simple estimate using Eq. (9) suggests that $0.999 T_c < T < T_c$ would be necessary to manifest the diverging simple quasiparticle relaxation length, which is beyond our experimental resolution for measurements of the corresponding $J_c$.

## V. DISCUSSION

The phenomenological analyses based on conventional theory of nonequilibrium superconductivity in the preceding section suggest significant effects of spin-injection on cuprate superconductivity and anisotropic spin transport, with spin relaxation probably dominated by the spin-orbit interaction along $c$ axis and by the exchange interaction within the $CuO_2$ plane. Under the premise of high-quality F-I-S heterostructures and interfaces, the significant influence of spin-polarized quasiparticles on the microscopic DOS is likely unique to the cuprate superconductors because of the strong correlation between the conducting holes and spin fluctuations.[3,9,10] Such drastic dynamic effects on cuprate superconductivity are reminiscent of the strong suppression of superconductivity and long-range effects induced by static nonmagnetic impurities that substitute the $Cu^{2+}$ ions in the $CuO_2$ planes.[17–28] The short-range antiferromagnetic correlation has been considered to play a significant role in the cuprate superconductivity, and the static nonmagnetic impurities in the $p$-type cuprates are believed to have broken the antiferromagnetic correlation of $Cu^{2+}$ ions,[16] thus inducing localized magnetic moments and resulting in suppression of the collective spin excitation and the global pairing potential.[21,24,25,27,28] Similarly, we consider that the continuous injection of spin-polarized quasiparticles into the cuprate superconductors has effectively resulted in a quasistatic ferromagnetic perturbation to the antiferromagnetic correlation in the $CuO_2$ planes, thereby yielding strong effects and slow relaxation in the quasiparticle spectra.

Next, we comment on the possible relevance of paramagnetic effect[80,84] to the observed suppression of cuprate superconductivity due to spin injection. We consider the spatially averaged effective magnetic induction $B_{eff}$ due to an injected spin-polarized current density $J_{inj}$. Assuming that the $c$-axis spin dephasing time $\tau_s^c$ and taking the polarization $P = 1$ for simplicity, we obtain an upper bound for $B_{eff}$:

$$B_{eff} \lesssim \mu_0 (\mu_B/e)(J_{inj} \tau_s^c/d), \quad (11)$$

where $\tau_s^c$ is related to $\tau_{so}$ as given in Eq. (10). Thus, we obtained $B_{eff} \sim 10^{-4}$ Tesla for $d = 100$ nm and $J_{inj} = 10^5$ A/cm$^2$. This effective field is clearly insignificant compared with any critical fields of the superconductor, thus cannot account for the strong suppression of superconductivity under the injection of spin-polarized quasiparticles.

For completeness, we discuss in the following the possibility that the suppression of critical currents might be a spurious effect associated with the summation of an initial supercurrent and an externally injected current, as recently suggested in Ref. 85, and then comment on the preliminary data taken on patterned F-I-S heterostructures. One may conjecture that as the externally injected current from the manganite underlayer enter the superconductor uniformly in a direction transverse to the $J_s$ measurement current in the superconductor, as shown in Fig. 6(a), this injected current would affect the measurements of $J_c^+$ and $J_c^-$ values differently due to the spatial variation in the local current density inside the superconductor. That is, one might assume that $J^+(y) = J_s + (y/L) J_{inj}$ and $J^-(y) = -J_s + (y/L) J_{inj}$, where $L$ is the length of YBCO along the $J_s$ direction, $\vec{J_s} \| \hat{y}$, and further conjecture that the $I$-$V$ characteristics of the entire superconductor would be solely determined by small resistive regions in the superconductor. More specifically, an apparent suppression of the measured $J_c$ would be expected because $J_c^+$ would be reduced by $J_{inj}$ due to its direct addition of $J_{inj}$ while $J_c^-$ would be unaffected and remains the same as $J_{c0}$.[85] However, upon closer scrutiny, we believe that such a hypothetical scenario has no merits for a number of reasons.

Empirically, all existing data derived from the as-grown and patterned devices can unambiguously rule out the current-summation scenario as the explanation for our observation. First, had the summing of currents as depicted in Fig. 6(a) been the dominating cause for the suppression in $J_c$, we would have found no change in $|J_c^-|$ and significant suppression only in $|J_c^+|$. However, such behavior has *never* been observed in any of our as-grown or patterned samples. Second, this two-dimensional geometrical effect would have resulted in a constant efficiency $\eta = 1/2$, for all $J_{inj}$ at all temperatures, and for all samples, regardless of the sample types (i.e., F-I-S vs N-I-S) and the constituent layer thickness. This clearly is contradicted by the experimental data shown in Figs. 5(a) and 5(b) for the as-grown heterostructures and in Fig. 6(c) on the patterned samples where $\eta$ varies significantly with $T$ and $J_{inj}$. In addition, for as-grown samples, no appreciable suppression in $J_c$ could be detected in the F-I-S devices with either a thicker superconducting layer ($d$ = 160 nm) [see Fig. 3(b)] or a thicker insulating barrier (10 nm) (Ref. 52) at all temperatures except very near $T_c^-$, implying $\eta \ll \frac{1}{2}$. Similarly, no discernible $J_c$ suppression could be found in the control N-I-S heterostructure with $d$ = 100 nm,[52] implying $\eta_n \sim 0$ for a wide range of temperature. Third, the simple current-summation scenario would assert that $J_c = 0$ if $J_{inj} \geq 2 J_{c0}$ for all heterostructures at all temperatures, which is at odds with the data shown in Fig. 1(c) for a control N-I-S sample. As mentioned previously in Sec. IV, such finding in the N-I-S samples is a clear revelation of the uncertainties in the injected quasiparticle momentum.[32] Fourth, we note that the experimental results by Vas'ko et al.[50] have demonstrated that the suppression $J_c$ in spin-injection devices is independent of the direction of current injection relative to the supercurrent, which further corroborate the notion that the positions of external electrodes do not provide well-defined supercurrent distribution within a superconductor. Finally, our previous scanning tunneling spectroscopic (STS) studies of the YBCO layer in both F-I-S and N-I-S heterostructures had demonstrated distinct changes in the quasiparticle DOS only under spin injection. The STS experiments were performed with $J_{inj} \geq 0$ and $J_s = 0$ at all times; hence, no complications from current addition were involved. Thus, we conclude that all experimental data to date clearly rule out the possibility of current summation as an alternative explanation for $J_c$ suppression in perovskite F-I-S devices.

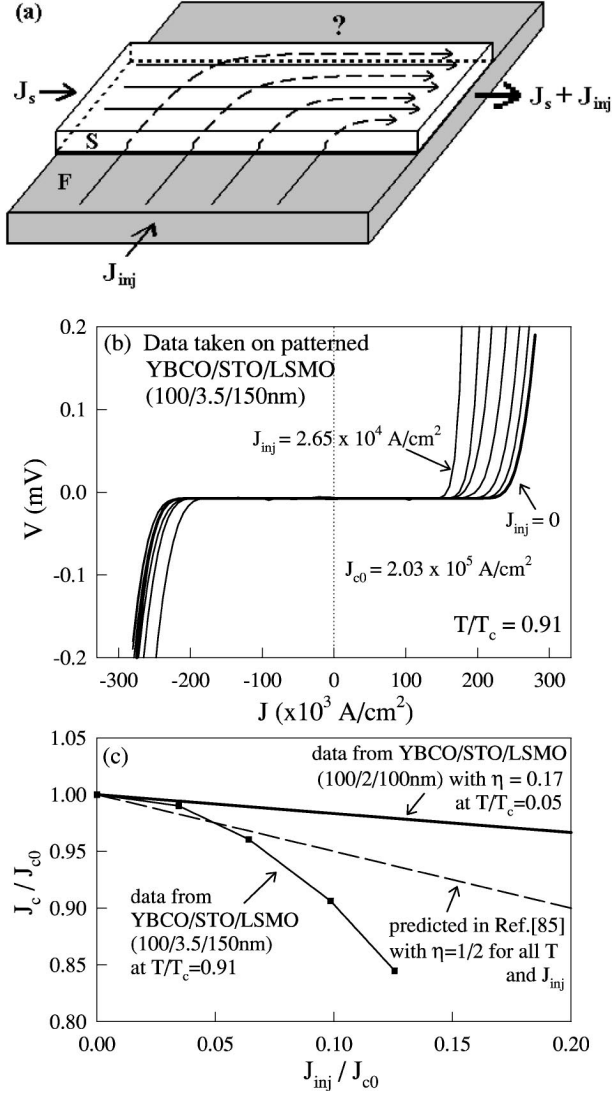

FIG. 6. (a) Hypothetical current flow patterns in the YBCO layer of the F-I-S heterostructure under external current injection from the manganite. The initial current in YBCO is $J_s$ along $\pm \hat{y}$ direction, and the external current enters the superconductor initially along the $\hat{x}$ direction. (b) I-V characteristics of a patterned F-I-S sample under injection currents from 0 to $2.65 \times 10^4$ A/cm$^2$. The lateral dimension of the superconductor is $(100 \times 100)$ $\mu$m$^2$, and the layer thicknesses are as indicated. (c) The dashed line represents the requisite observation of efficiency $\eta$ equaling a constant 1/2, if the geometrical effect of current summation in (a) were correct. The lower solid line is the result derived from (b), with the efficiency $\eta$ varying continuously with $J_{inj}$. The upper solid line is derived from another patterned F-I-S sample of $(10 \times 100)$ $\mu$m$^2$ lateral dimension and thicknesses as indicated, with $\eta = 0.17$. Clearly, the data do not support the current-summation scenario proposed in (a).

From the theoretical viewpoints, the current-summation scenario assumes that the injected quasiparticles follow a well-defined current path, which immediately turn after entering the superconductor, flow toward the common-ground terminal, and exit the superconductor after aggregating at that end of the superconductor, as depicted in Fig. 6(a). In other words, although the incident quasiparticle momentum was along the $c$ axis, only the final momentum parallel to the direction of the supercurrent in the CuO$_2$ planes was considered relevant. Such approach is unjustified for the following reasons. First, the hypothetical geometric effect for a partially varying total current density in the superconductor would have resulted in a phase gradient in the order parameter throughout the superconductor. Such a gradient would have incurred phase slippage and vortex formation in the superconductor, and the interaction of the nonuniform currents with vortices would tend to redistribute the currents more uniformly to minimize the phase gradient. Thus, the real current distribution inside the superconductor is expected to deviate from the direct sum of currents as depicted in Fig. 6(a). Second, the dynamic nature associated with the initial interaction of the injected quasiparticles with the superconductor plays a very important role in determining the nonequilibrium superconducting properties, such as the overall quasiparticle energy and the DOS. These important processes such as the quasiparticle redistribution and pair recombination could not be neglected unless the quasiparticle relaxation times were sufficiently short so that the corresponding characteristic lengths were much smaller than the sample dimensions. However, as we have estimated in Sec. IV, the in-plane spin-relaxation time could range from $10^{-4}$ s at $(T/T_c) \ll 1$ to $10^{-6}$ s at $(T/T_c) \to 1$, so that the in-plane spin-relaxation length was on the order of $10^{-4}$–$10^{-3}$ m, comparable to the device dimension. The nonequilibrium effect of spin-polarized quasiparticles appeared to be long range at all temperatures in F-I-S samples with thin YBCO, and therefore cannot be neglected. Third, the current-summation scenario ignores the dependence of quasiparticle transmission across interfaces on the degree of spin polarization, the quasiparticle energy, the no double occupancy constraint, and the interface properties. Such simplification is neither theoretically rigorous nor empirically compatible with experimental data.

In an earlier study of an N-I-S heterostructure,[86,87] a suppression of $J_c^+$ obtained was attributed to the effect of current summation in Ref. 85. Upon closer inspection of the experiments, it can be ascertained that the critical current had been determined only in one supercurrent direction ($J_c^+$), while the reversal of polarity was done to the injection current in the measurements. In reality, the experimental procedure in Ref. 87 gave rise to a branch imbalance effect associated with the injection of charged quasiparticles.[32] That is, reversing the polarity of the injection gate current actually changed the injected quasiparticles from predominately electronlike to holelike in character, or vice versa. It is known that a complete description for the branch imbalance effect must include studies of both polarities of the bias voltage.[32] Similarly, showing the full $J_c$ behavior requires the mapping of both $J_c^+$ and $J_c^-$ values, but no results for $J_c^-$ were reported in Ref. 87. Therefore, the current summation conjecture remains unsubstantiated, contrary to the assertion in Ref. 85.

We note, however, that the effect of $J_c$ suppression due to spin injection is weaker in the patterned F-I-S relative to the as-grown F-I-S devices, as shown in Fig. 6. In particular, for the patterned F-I-S samples of thicknesses 100 nm/2 nm/100 nm, we find that $\eta \approx 0.17$ at $T = 4.2$ K, while at higher temperatures, no discernible $J_c$ suppression (i.e., $\eta \approx 0$) is observed even with $J_{inj} > 2 J_c$. As described in Sec. II, the constituent layers of most patterned F-I-S heterostructures have shown substantial degradation particularly near the edge of the YBCO layer. Consequently, the degree of spin polarization is likely to be significantly compromised. Furthermore, severe interface magnetic scattering becomes likely as the result of overall material degradation. Thus, the weaker spin-injection effect on the patterned F-I-S devices is not conclusive, and should not be considered as inconsistent with our estimated long in-plane spin-relaxation length. However, ultimate empirical verification for the in-plane spin-relaxation length awaits successful fabrication of high-quality patterned F-I-S and N-I-S devices.

On the magnitude of the efficiencies $\eta$ associated with both F-I-S and N-I-S heterostructures, we note that they are generally small except at low temperatures in the F-I-S. This is not entirely surprising because the YBCO superconductor is known to have $d$-wave pairing symmetry, which is gapless along the nodal directions. The pre-existence of thermally excited quasiparticles diminishes the significance of those injected externally. Only in the low temperature regime, where the nonequilibrium effects become significant, does one observe larger $\eta$ in the spin-polarized quasiparticle injection. Similar findings of small efficiencies under injection have been confirmed by a different experimental technique through magnetization measurements of YBCO films.[88]

Finally, based on the phenomenological analysis of our experimental data outlined in this paper, we remark that the bulk nonequilibrium effects in perovskite F-I-S and N-I-S heterostructures appear to be conceptually consistent with the general descriptions for quasiparticles. In other words, there is no obvious need to invoke spin-charge separation in the superconducting state to account for the spin and charge transport behavior in the cuprates.

## VI. CONCLUSION

We have conducted systematic studies of the critical current density ($J_c$) in perovskite F-I-S and N-I-S heterostructure with different thicknesses of the superconducting layer, and have demonstrated sharp contrasts between the temperature and injection current dependence of F-I-S and of N-I-S. Within experimental uncertainties, the strong suppression of superconductivity in F-I-S due to current injection cannot be trivially explained by either the paramagnetic effect or a simple current-summation effect. Phenomenological analyses of our data suggest that the strong influence of spin-polarized quasiparticles on $J_c$ and on the quasiparticle density of states of F-I-S samples may be due to their suppression of the antiferromagnetic correlation in the $CuO_2$ planes of the superconducting cuprate. Assuming the applicability of conventional theory of nonequilibrium superconductivity, the strong effects of spin-polarized quasiparticles are manifested by the long in-plane spin-relaxation time and large shift in the chemical potential derived herewith. In contrast, no discernible chemical potential shift can be found in the N-I-S samples using the same analysis. The strong effects of spin-polarized quasiparticles are probably unique to the cuprates and other superconductors that exhibit coexistence of antiferromagnetic correlation and superconductivity, and are reminiscent of the significant suppression of superconductivity due to nonmagnetic impurities in the $CuO_2$ planes. In contrast to the in-plane spin relaxation mechanism via exchange interaction, the $c$-axis spin-transport mechanism may be dominated by inelastic spin-orbit interaction. Although more accurate determination for the spin-relaxation times awaits successful fabrication of patterned devices with well-defined geometry and high-quality interfaces, our work has demonstrated phenomena of nonequilibrium superconductivity in cuprate superconductors and the strong effects of spin injection. Further theoretical studies for the microscopic interaction of spin-polarized quasiparticles with the background antiferromagnetic correlation in the highly anisotropic $d$-wave cuprates will be necessary to provide better understanding of the data.

## ACKNOWLEDGMENTS

The work at Caltech was jointly supported by NSF Grant No. DMR-0103045 and NASA/OSS. We are grateful to Dr. R. P. Vasquez at the Jet Propulsion Laboratory for providing the F-I-S and N-I-S heterostructures, performing the XPS characterization, and for useful discussions.